\documentclass[conference]{IEEEtran}

\usepackage{graphicx}
\usepackage{amsmath}
\usepackage{relsize}
\usepackage{booktabs}
\usepackage{subcaption}
\usepackage{amsmath}
\usepackage{algorithm}
\usepackage{algorithmic}
\usepackage{color}

\begin{document}
\title{Combination of Visible Light and Radio Frequency Bands for Device-to-Device Communication}
\author
{\IEEEauthorblockN{Pavel Mach, Zdenek Becvar, Mehyar Najla, Stanislav Zvanovec}
\IEEEauthorblockA{
Faculty of Electrical Engineering, Czech Technical University in Prague\\
Technicka 2, 166 27 Prague, Czech republic\\
emails: \{machp2, zdenek.becvar, najlameh, xzvanove\}@fel.cvut.cz}
}

\maketitle
\begin{abstract}
Future mobile networks are supposed to serve high data rates required by users. To accommodate the high data rates, a direct communication between nearby mobile terminals (MTs), known as Device-to-device (D2D) communication, can be exploited. 
Furthermore, a communication in high frequency bands, such as, visible light communication (VLC), is also foreseen as an enabler for the high data rates. 
In conventional D2D communication, pairs of the MTs should reuse the same frequencies to keep a high spectral efficiency of the system. 
However, this implies either interference among the D2D pairs or utilization of complex and advanced resource allocation algorithms. 
In this paper, we propose a new concept for D2D communication combining VLC and RF technologies in order to maximize capacity of the system. 
This paper provides  an analysis of the operational limits for the proposed concept and investigates capacity gains introduced by the combined usage of RF and VLC bands for D2D. 
Moreover, we discuss several practical issues related to the proposed RF--VLC D2D concept. 
Performed analyses show that the RF--VLC D2D is able to improve the capacity in an indoor scenario by a factor of 4.1 and 1.5 when compared to stand-alone RF D2D and VLC D2D, respectively.  
\end{abstract}

\section{Introduction}
In conventional mobile networks, mobile terminals (MTs) communicate through a base station, in LTE-A denoted as an evolved node B (eNB). The concept of a direct communication among the MTs in proximity of each other, known as Device-to-Device (D2D) communication, is considered as a way to enhance the capacity of mobile networks and to increase spectral efficiency \cite{3gpp36.877}. Furthermore, D2D enables to decrease a packet delay and to reduce a power consumption of the MTs due to mutual proximity of the MTs \cite{mach2015band}.

In general, D2D communication can be used in either in-band or out-band fashions. In the case of in-band D2D, the MTs reuse the same frequency bands as a conventional cellular communication, e.g., licensed frequencies allocated for mobile networks. Hence, interference between D2D and the conventional cellular communication is seen as a critical problem \cite{fodor2012design}. To address this problem, many interference mitigation techniques, such as power control \cite{xu2014dynamic}, radio resource allocation \cite{zhang2013interference}, scheduling \cite{wang2013joint}, etc., can be applied. Nonetheless, if interference between the D2D and cellular communications cannot be sufficiently mitigated by these techniques, D2D may be forced to operate in a dedicated mode (also known as an overlay mode). In the dedicated mode, D2D exploits orthogonal resources to the cellular communication to avoid interference entirely, however, it is at the cost of decreased spectral efficiency \cite{zulhasnine2010efficient}.

In the case of the out-band D2D, the D2D communication takes place in unlicensed bands through WiFi-Direct or Bluetooth, as investigated, e.g., in \cite{lin2015hybrid}. Nevertheless, if D2D pairs in close vicinity of each other reuse the same out-band frequencies, interference among D2D pairs remains a problem and limits the benefits of D2D. To minimize interference among D2D pairs, visible light communication (VLC) can be also considered for the out-band D2D. The VLC systems operate at wavelengths of 380-750 nm (i.e., frequency bands of 400-790 THz) \cite{jungnickel2015european} and can achieve high data rates. For example, 4.5 Gbps throughput can be achieved by the VLC systems employing carrier-less amplitude \& phase modulations and a recursive least square-based adaptive equalizer as described in \cite{Standa1} and \cite{Standa2}, respectively. In \cite{Standa3}, the authors show that a combination of 16--quadrature amplitude modulation and orthogonal frequency division multiplexing (OFDM) or wavelength multiplex (RGB) allow to reach 3.4 Gbps throughput. A disadvantage of VLC can be seen in a low scalability for longer distances and its susceptibility to a volatility of the MT's orientation resulting in sudden decrease in channel quality even for small changes of the MTs' orientation (in terms of irradiance and incidence angles) \cite{pathak2015visible}.

A combination of communication in the conventional radio frequency (RF) and VLC bands is investigated, e.g., in \cite{li2015joint} and \cite{liang2015novel}. In both studies, the authors assume that the VLC access points are deployed at the ceiling and that RF and VLC bands are used for uplink and downlink communication, respectively. Nevertheless, these papers do not consider D2D communication, which introduces new challenges and opportunities related to a higher volatility of both sides of the communication chain and proximity of the MTs. To our best knowledge, the VLC for D2D is considered only in \cite{liu2016game} and \cite{tiwari10optical}. In \cite{liu2016game}, the authors propose a game theory-based mechanism choosing the optimal mode of VLC communication from three candidate modes in order to enhance channel capacity. The first mode is a direct VLC communication (VLC D2D), the second mode is a indirect VLC communication through an access point and the third mode represents a mix of the first two modes. In other words, the paper investigates behavior of a conventional D2D in VLC bands. In \cite{tiwari10optical}, an optical repeater-assisted VLC D2D system is presented. The VLC repeater enables VLC for longer distances and allows to enhance VLC range when the direct link between the MTs is not available. This is an analogy to D2D relaying as addressed frequently in the conventional D2D in RF bands. However, even \cite{tiwari10optical} is focused purely on VLC bands and does not consider a combination of RF and VLC for D2D. 

In this paper, we propose a new concept combining in-band RF and out-band VLC for D2D communication. The proposed concept takes advantage of the fact that RF and VLC do not interfere to each other and VLC signal is strongly attenuated with distance, thus, interference to other D2D pairs operating in VLC is naturally suppressed. At the same time, RF enables to preserve benefits of common D2D for larger distances at which VLC cannot operate. By allowing selection of either RF or VLC for each D2D pair, overall level of interference is significantly reduced and the system capacity is increased. To motivate further research in the area of combined RF-VLC D2D, we discuss an applicability of the new concept and contemplate key practical issues in order to implement RF-VLC D2D. Then, we investigate limits of the operation and gains introduced by RF-VLC D2D depending on various parameters, such as a distance between the MTs of the same D2D pairs, a distance between D2D pairs, or irradiance/incidence angles of the MTs. As this paper is an initial work in this domain, we limit our investigation to indoor scenario where we foresee main benefits of the VLC-RF D2D. Moreover, we assume that the RF D2D uses dedicated resources with respect to the conventional cellular communication and, thus, only interference among D2D pairs exploiting RF is an issue. Through simulations, we show that the combination of RF and VLC for D2D allows a significant increase in the capacity of the system.  

The paper is organized as follows. In Section II, a system model for the proposed RF-VLC D2D concept is defined. In Section III, we contemplate key practical issues for D2D combining RF and VLC bands and we discuss potential use-cases for the proposed concept. Then, Section IV is dedicated to a description of the simulation scenarios and to a discussion of the simulation results. The last section concludes the paper and outlines future research directions.

\section{System model for RF--VLC D2D}
\begin{figure}[!t]
\centering
  \includegraphics[scale=0.75]{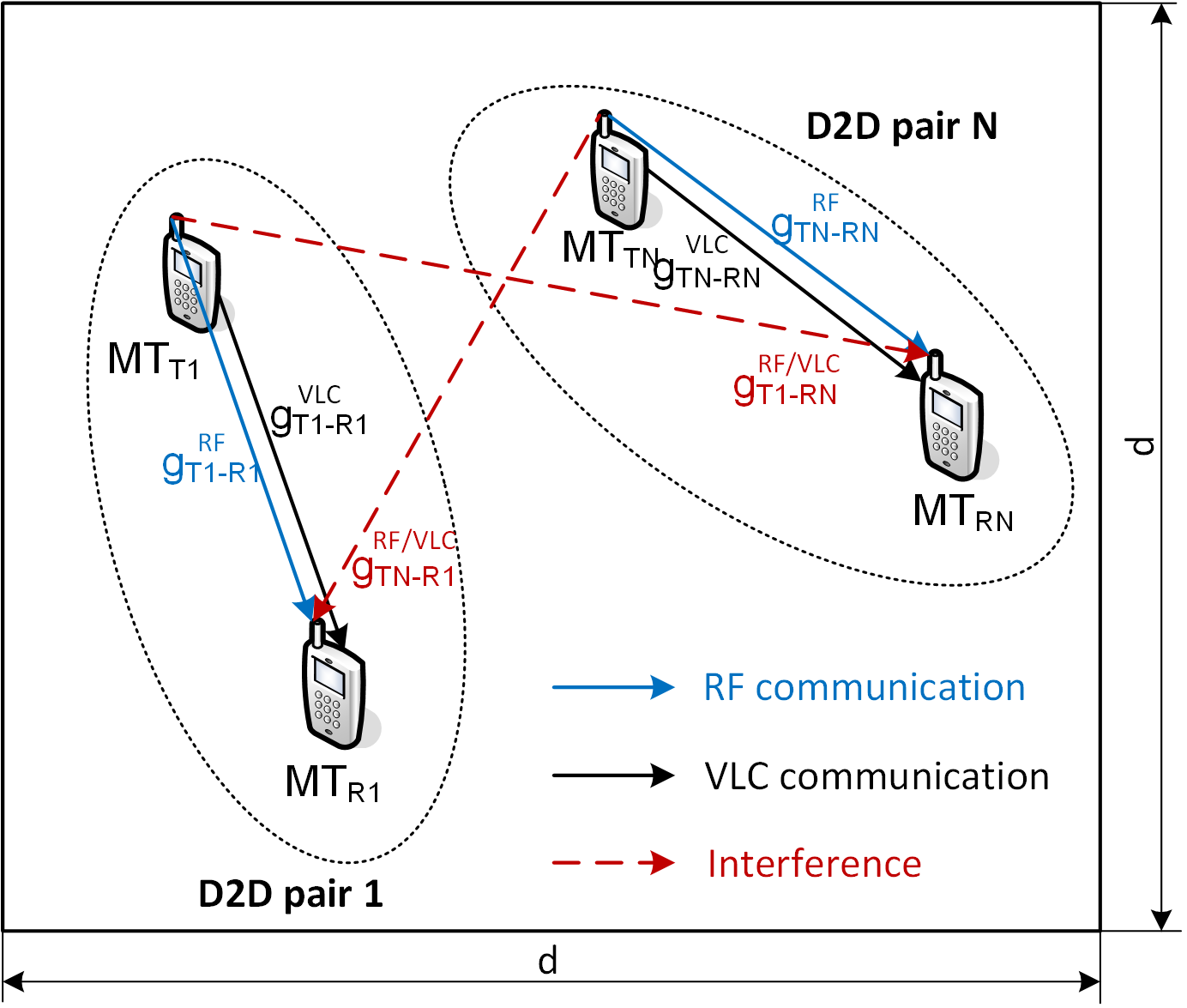}
  \caption{System model for investigation of the proposed RF-VLC concept.}
  \label{Fig01}
\end{figure}

In this section, we describe a general system model and mode selection for the proposed RF-VLC D2D concept. We assume $N$ MTs randomly distributed within a square room with a dimension $d$ (see Fig~\ref{Fig01}). As VLC is highly susceptible even to small changes in the angles between a transmitting MT ($MT_T$) and a receiving MT ($MT_R$) \cite{Standa}, we assume varying azimuthal orientation of both MTs. Note that the varying angles are more critical in terms of VLC channel quality than the MTs' mobility, since the mobility leads to a continuous and slow changes of the angles between the $MT_T$ and the $MT_R$. In contrast, turning the MTs leads to an immediate and a steep change of the angles. Thus, for sake of simplicity and clarity, we leave the mobility of the MTs for further research and we focus on static MTs in this paper.

Among all $N$ active MTs, $N_p$ D2D pairs are randomly selected so that every MT is involved in just one D2D pair (i.e., $N_p=N/2$). The channel gain between the MTs within one D2D pair is denoted as $g_{T-R}^{RF}$ and $g_{T-R}^{VLC}$ for RF and VLC modes, respectively. We assume that the D2D pairs exploit dedicated uplink resources with respect to the cellular communication so there is no interference between the D2D MTs and MTs communicating with the eNBs. Contrary, all D2D pairs operate in the same RF bands and, thus, interfere with each other (see Fig.~\ref{Fig01} where the $MT_{T1}$ causes interference to the $MT_{RN}$ and the $MT_{R1}$ is interfered by the $MT_{TN}$). Consequently, the available capacity for RF D2D is significantly influenced by the amount of interference originating from nearby D2D pairs. The $MT_T$ exploiting VLC does not introduce interference to the $MT_R$ operating in RF as these operate at different frequencies. 

The communication mode (either RF or VLC) is selected with objective to maximize the system capacity, that is:
\begin{equation}
\mathcal{M} = \begin{cases}
             RF:  & C^{RF} \geq C^{VLC}  \\
             VLC: & C^{RF} < C^{VLC} 
       \end{cases} \quad
\end{equation}
where capacities of both RF D2D ($C^{RF}$) and VLC D2D ($C^{VLC}$) are derived according to Shannon-Hartley’s theorem:
\begin{equation}
C^{RF}=B^{RF}{log_{2}(1+SINR^{RF})}
\end{equation}
\begin{equation}
C^{VLC}=B^{VLC}{log_{2}(1+SINR^{VLC})}
\end{equation}
where $B^{RF}$($B^{VLC}$) is the system bandwidth of RF (VLC), $SINR^{RF}$ stands for the signal to interference plus noise ratio (SINR) observed by the $MT_R$ in RF mode, and $SINR^{VLC}$ represents SINR experienced by $MT_R$ in VLC mode. The $SINR_{Rn}^{RF}$ experienced by the $n$-th MT ($MT_{Rn}$) is expressed as:

\begin{equation}
SINR_{Rn}^{RF}=\frac{P^{RF}_{t } {g_{Tn-Rn}^{RF}}}
{\sum_{i\neq n}(P_{t}^{RF}{g_{Ti-Ri}^{RF})}+\sigma^2_{t, RF}}
\end{equation}
where $P^{RF}_t$ is the RF transmitting power of the $MT_T$, $g_{Ti-Ri}^{RF}$ is the RF channel gain between the $MT_{Ti}$ and the $MT_{Ri}$ of the $i$-th D2D pair, and $\sigma^2_{t, RF}$ stands for the thermal noise in RF with a spectral density of --174 dBm/Hz. 

The $SINR_{Rn}^{VLC}$ experienced by the $MT_{Rn}$ is defined as:

\begin{equation}
SINR_{Rn}^{VLC}=\frac{P^{VLC}_tg^{VLC}_{Tn-Rn}\gamma^2}
{\sum_{i\neq n}(P_{t}^{VLC}{g_{Ti-Ri}^{VLC})}+\sigma^2_{t, VLC}+\sigma^2_{s}}
\end{equation}
where $P^{VLC}_t$ represents the transmitting optical power of the transmitting LED, $g^{VLC}_{Tn-Rn}$ is the VLC channel gain between the MTs of the $n$-th D2D pair, $\gamma$ is the responsivity of a photo-diode, and $\sigma^2_{s}$ corresponds to the shot noise. 

The VLC channel gain $g^{VLC}_{Tn-Rn}$ is strongly dependent on the irradiance angle ($\phi$), incidence angle ($\psi$), and on the parameters of the optical receiver. Thus, the channel gain $g^{VLC}_{Tn-Rn}$ is expressed by the following equation:
\begin{equation}
g^{VLC}_{Tn-Rn}=\frac{(m+1)A{cos^m(\phi)}T_sg(\psi)cos(\psi)}
{2\pi d_{TR}^2}
\end{equation}
where $A$ is the physical area of a photodetector, $T_s$ stands for the gain of an optical filter, $d_{TR}$ is the distance between the $MT_T$ and $MT_R$, $g$(\(\psi\)) is the gain of an optical concentrator, and $m$ corresponds to the order of Lambertian emission defined as follows:
\begin{equation}
m=\frac{-ln(2)}
{ln(cos(\phi_c))}
\end{equation}
where $\phi_c$ is the transmitter semi-angle at half power \cite{pathak2015visible}. The gain of the optical concentrator ($g$(\(\psi\))) depends on the photodetector view angle ($\psi_c$) and it is expressed as:
\begin{equation}
g(\psi) = \begin{cases}
             \frac{n^2}{sin^2(\psi_c)} & \text{if } 0<\psi \leq \psi_c \\
             0  & \text{} otherwise
       \end{cases} \quad
\end{equation}
The thermal and shot noises for VLC are calculated as:
\begin{align}
\begin{split}\\\sigma^2_{t, VLC}=(\frac{8\pi kT_k\eta AI_2B^2}{G}) ~+ \\ ~+ (\frac{16\pi^2kT_k\Gamma\eta^2A^2I_3B^3}{g_m})
\end{split}
\end{align}
\normalsize
\begin{equation}
\sigma^2_{s}=(2qI_{bg}I_2B)+(2q\gamma P_t^{VLC} g^{VLC}_{Tn-Rn}B)
\end{equation}
where $k$ is Boltzmann's constant, $T_{k}$ corresponds to the absolute temperature, $\eta$ is the fixed capacitance of the photodetector per unit area, $I_2$ and $I_3$ stand for the noise bandwidth factors, $B$ represents the equivalent noise bandwidth, $G$ is the open-loop voltage gain, $\Gamma$ is FET channel noise factor, $g_m$ corresponds to FET transconductance, $q$ is the charge, and $I_{bg}$ is the background current \cite{pathak2015visible}.
We assume the MTs are equipped with the RGB-based LED and the photodetector at the transmitter and the receiver, respectively \cite{tiwari10optical}.

\section{Practical issues and use cases for RF-VLC D2D}
In this paper, we demonstrate a potential efficiency of the combined RF-VLC D2D. Nevertheless, there are several practical issues that have to be contemplated in order to bring the whole concept into fruition. Hence, this section discusses the applicability of the RF-VLC D2D in real network and also discusses some practical issues of the proposed concept. 

The first important aspect regarding the combination of RF and VLC for D2D is to outline its use-cases and suitable scenarios. Basically, we can expect that the combination of RF and VLC for D2D would be beneficial for future services requiring high throughput and low latencies. In general, low throughput services or calls are not seen as the most promising options for the RF-VLC D2D due to their demands on relatively low capacity and high sensitivity to sudden connection degradation. Thus, rather services and applications requiring a high throughput while tolerating a throughput variation are supposed to be good target for the RF-VLC D2D. Then, since the VLC is beneficial especially for short distances, we can expect that RF-VLC D2D concept should be used indoor, where users who want to transmit a high amount of data to another users (e.g., exchanging photos or videos) can direct their MTs towards each other in order to enhance capacity by VLC. To this end, we analyze requirements on angles between $MT_T$ and $MT_R$ later in this paper.

The second important aspect regarding the combination of RF and VLC for D2D is to decide whether the control signaling can be transmitted also in both transmission modes (RF/VLC), like data transmission, or not. Although VLC may offer superior capacity for short distances when compared to RF, this is true only for optimal or near optimal irradiance and incidence angles (as further discuss in the next section). As a matter of fact, the VLC channel is highly susceptible to these changes and, hence, sudden decrease in channel quality may occur frequently. Consequently, the control signaling must be unconditionally transmitted via RF D2D link during the whole communication. This is supported by the fact that the amount of the signaling is incomparably lower than the amount of user’s data and the capacity offered by VLC cannot be fully exploited for the signaling anyway.

Another important question is: who decides which communication mode (RF or VLC) should be selected for data transmission at the moment? In general, D2D communication may be controlled in a centralized or a distributed manner. In the former case, the selection is done solely by the eNB. Consequently, the MTs have to report the information regarding the channel quality to the eNB on regular basis. Since the channel quality may vary significantly, especially for VLC channel, the delay in decision at the eNB may result in an incorrect selection of the communication mode resulting even in a capacity degradation. In the latter case, if the selection is performed directly at the MTs (i.e., in the distributed manner), the delay of the decision is significantly reduced. In general, the mode selection can be carried out by both, the MT$_T$ and the MT$_R$. Nonetheless, we suggest to make the decision at the MT$_T$ rather than at the MT$_R$ as the MT$_T$ is aware of the transmission buffer status and can perform scheduling of the RF and VLC resources accordingly. To this end, the quality of both RF and VLC channels has to be reported by the MT$_R$ to the MT$_T$ via RF. Then, the MT$_T$ can promptly react to rapid degradation of VLC channel quality and switch to RF for data transmission immediately. In fact, the single scheduler can serve both VLC and RF communication without any complication as the schedulers perceive both technologies from a perspective of scheduling metrics (capacity, delay, buffer status, etc.), which can be represented in the same way for both technologies. Therefore, there is no need for any advanced mechanisms to control the proposed RF-VLC D2D concept.

\section{Performance evaluation}
In this section, the simulation scenarios and main simulation parameters are described. Then, results of the simulations for individual scenarios are outlined and discussed.

\subsection{Simulation scenario and models}
We first outline scenario considered for the performance evaluation and comparison of the proposed RF-VLC D2D with competitive solutions. Then, we also define models and parameters considered in the simulations.

\begin{figure}[b!]
\centering
  \includegraphics[scale=0.47]{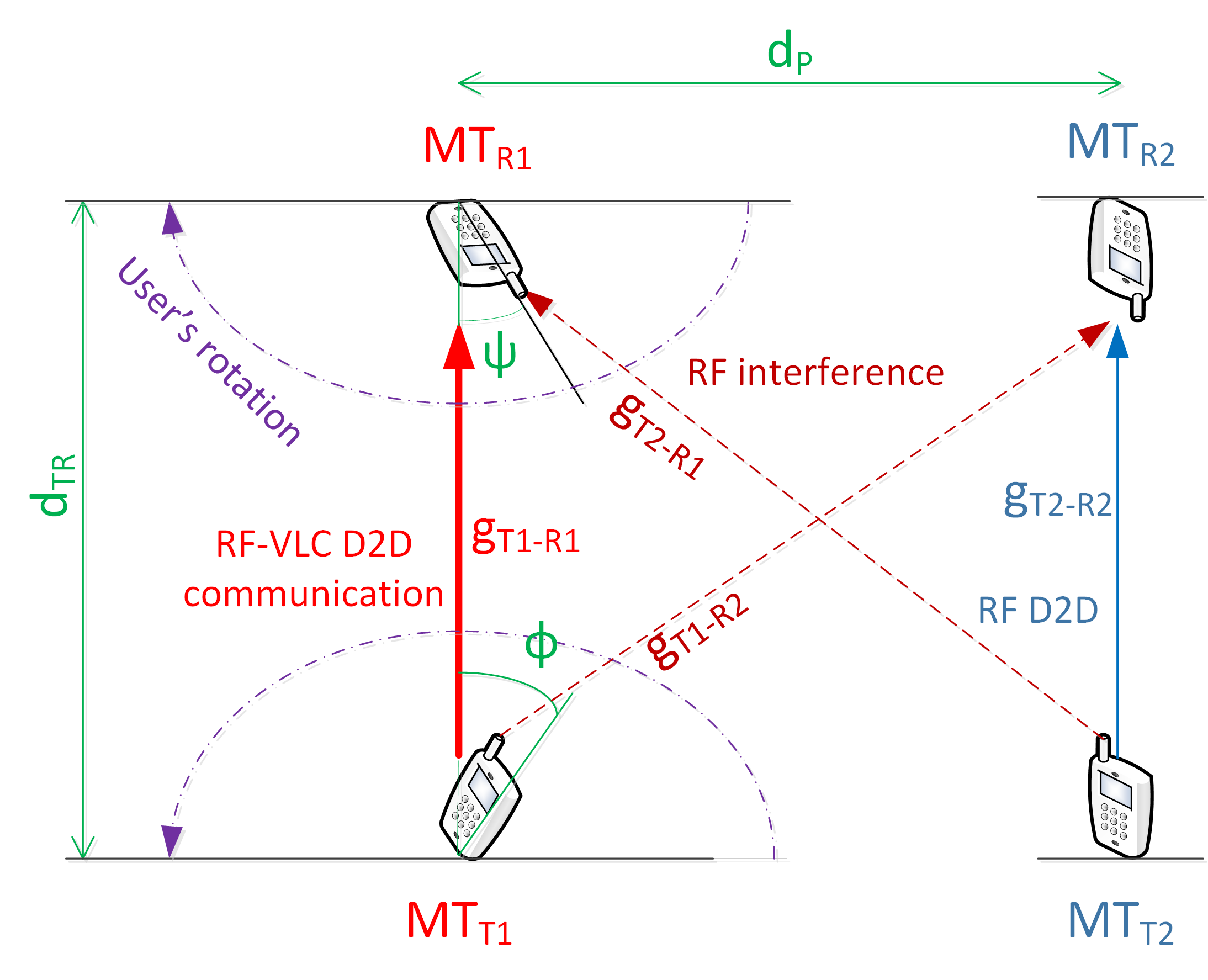}
  \caption{Explanation of parameters and deployment scenario considered for performance assessment.}
\label{Fig02}
\end{figure}

We assume a scenario representing an indoor area (a room or a hall) where we foresee main benefits of the proposed concept as explained in Section III. Further, we assume the room dimension of $d$ x $d$ m. In the room, four MTs are deployed within specific distance of $MT_T$ and $MT_R$ of the same pair ($d_{TR}$) and with specific inter-pair distance $d_P$ as shown in Fig~\ref{Fig02}. We set these distances manually to understand behavior of the RF--VLC D2D over various distances in order to assess potential limits and scalability of the solution.

\begin{table}[t!]
	\centering
	\caption{Simulation parameters.}
	\label{tab:table1}

\begin{tabular}{ |ll|l| }
\hline
\multicolumn{3}{ |c| }{RF Parameters} \\
\hline
\multicolumn{2}{ |c| }{ Parameters}  & Value  \\ \hline
Carrier frequency & $f_{c}$ & 2 [MHz]\\
Bandwidth & $B^{RF}$ & 20 [MHz] \\
Transmission power of MT &$P_t^{RF}$ & 200 [mW] \\
\hline
\multicolumn{3}{ |c| }{VLC Parameters} \\
\hline
\multicolumn{2}{ |c| }{ Parameters}  & Value  \\ \hline
Bandwidth& $B^{VLC}$ & 10 [MHz] \\
Transmission power of MT& $P_t^{VLC}$ & 200 [mW] \\
Physical area of photodetector &$A$& 1 ${[cm]^{2}}$ \\
Background current		&\(I_{bg}\)  &     10 [nA] \\
Noise Bandwidth factors &\(I_2\)-\(I_3\) &     0.562  -   0.0868 \\
Fixed capacitance of the photodetector& \(\eta\) &      112$\times{10^{-8}}$  [F/m]\\
FET channel noise factor		&\(\Gamma\) &     1.5 \\
FET transconductance&	\(g_m\)  &     0.03 [s] \\
		Responsivity of the photo diode & \( \gamma \)&    0.53 [A]  \\
		Open-loop voltage gain & \(G\)&     10 \\
	Optical concentrator gain	& \(g\)(\(\psi\))  &3      \\
	Optical filter gain	& \(T_s\)&      1\\
Absolute temperature	&	$T_{k}$&      295 [k]\\
\hline
\multicolumn{3}{ |c| }{General Parameters} \\
\hline
\multicolumn{2}{ |c| }{ Parameters}  & Value  \\ \hline
Number of MTs  & $N$ & 4 \\
Irradiance angle & $\phi$  &  -90 -- 90 [$^{\circ}$] \\
Incidence angle & $\psi$  &  -90 -- 90 [$^{\circ}$] \\
Room dimension &$d$  &  30 [m] \\
\hline
\end{tabular}
\label{Tab01}
\end{table}

The orientation (azimuth) of each MT is generated in one of the following ways: Optimal, Gaussian, and Random selection. The Optimal selection means that the $MT_T$ and $MT_R$ are oriented directly towards each other (i.e., in Fig.~\ref{Fig02}, $\phi$ and $\psi$ are set to 0$^{\circ}$). This case shows an upper bound performance as it allows reaching maximum capacity for VLC mode. In the case of Gaussian selection, the $\phi$ and $\psi$ angles are randomly generated according to the Gaussian distribution with the mean ($\mu$) set to 0$^{\circ}$ and the standard deviation ($\sigma$) set to 60$^{\circ}$. This situation represents the case when two users are willing to exchange data and are aware of each other locations so that we assume they try to direct their MTs towards each other. Nevertheless, even if the users try to direct their MTs towards each other, they might not match the angles in a perfect way so there is a possibility of a deviation from the optimal angles. The third option, Random selection, shows one of the worst cases since $\phi$ and $\psi$ angles are selected randomly between 0$^{\circ}$ to 180$^{\circ}$. This situation can appear when both users cannot or do not want to change orientation of their MTs and keep the MT in a random direction with respect to the other MT.

The $MT_T$ transmit data to the $MT_R$ in a mode (RF or VLC) which provides higher capacity at the moment as described in the system model (Section II). For the RF channel, we follow the channel modeling defined by 3GPP for indoor D2D communication as defined in \cite{stefan2013area}. The modeling of VLC channel is performed according to \cite{pathak2015visible}. Detailed parameters of both channels and general simulation parameters are summarized in Table~\ref{Tab01}.
 
\begin{figure*}[t!]
\centering
\begin{subfigure}[t]{0.5\textwidth}
\centering
\includegraphics[width=\textwidth]{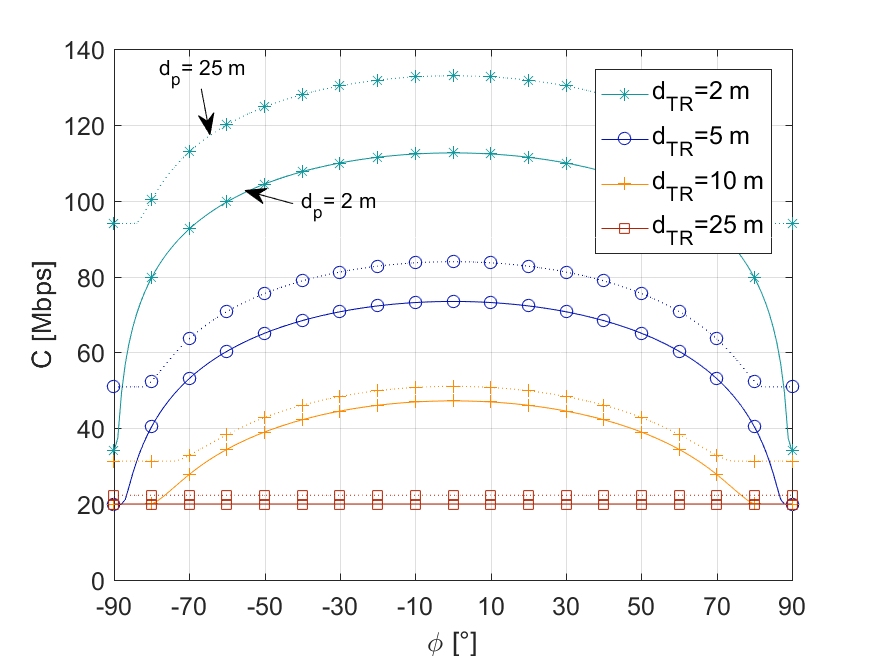}
\caption{}
\end{subfigure}%
~
\begin{subfigure}[t]{0.5\textwidth}
\centering
\includegraphics[width=\textwidth]{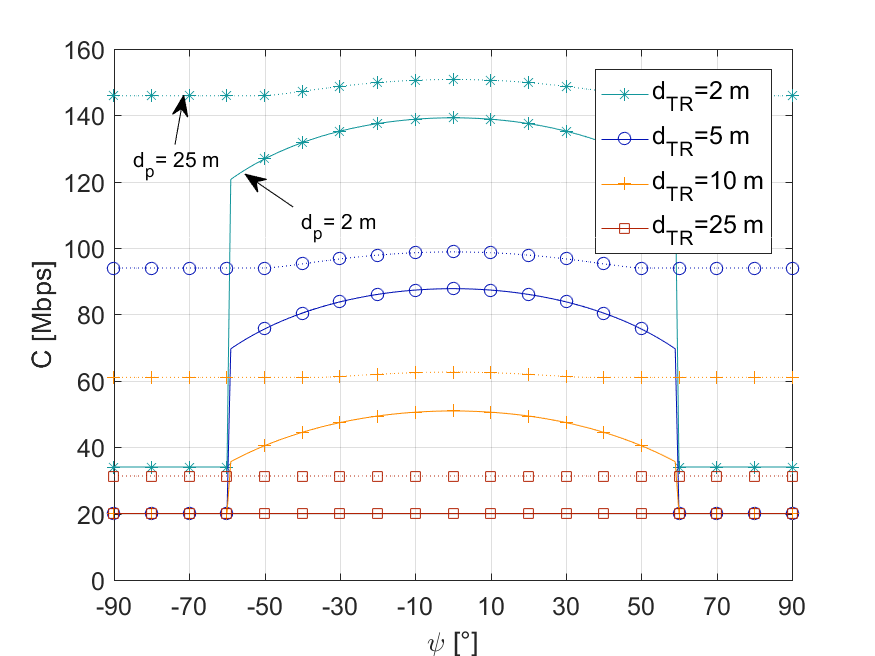}
\caption{}
\end{subfigure}
\caption{Performance of RF-VLC D2D for various angles $\phi$ (subplot (a) and $\psi$ (subplot (b)) with the capacity averaged out over 180 values of the second angle (i.e., $\psi$ in subplot a) and $\phi$ in subplot b)) ranging from -90${^\circ}$ to 90${^\circ}$ with a step of 1${^\circ}$ (solid lines represent $d_p = 2$~ m, dotted lines represent $d_p = 25$~m).}
\label{fig:1a}
\end{figure*}

\subsection{Simulation results and discussion}
In this section, we present the results obtained by the simulations. First, we analyze an impact of the angles $\psi$ and $\phi$ on the D2D capacity to understand the potential performance gain of RF-VLC D2D. Second, we investigate an impact of the $d_{TR}$, $d_{P}$, and $\phi$ on VLC usage ratio (i.e., how often VLC is used instead of RF). Third, we compare the capacity achieved by the proposed RF-VLC D2D system with RF D2D (i.e., without VLC) and VLC D2D (without RF). 

Fig.~\ref{fig:1a} demonstrates the impact of angles on the capacity achieved by the proposed RF-VLC D2D system for various $d_{TR}$ and $d_{P}$ distances . For irradiance angle ($\phi$), the change in capacity is continuous as the LED diode can operate in the whole range of 90${^\circ}$ while for incidence angle ($\psi$), the communication is limited by the field of view (FOV) of the photodetector (set to 60${^\circ}$ in this paper according to \cite{burton2012performance}). The $\phi$ ($\psi$) is set gradually from -90${^\circ}$ to 90${^\circ}$ in respective figures. For each angle, the capacity is computed as an average value achieved over corresponding $\psi$ ($\phi$) ranging from -90${^\circ}$ to 90${^\circ}$ with a step of 1${^\circ}$. We can see that the $d_{TR}$ distance plays a crucial role in the capacity. For smaller distances, i.e., if $d_{TR}<10$ m, the capacity rises significantly with increasing $d_{p}$. For a larger $d_{TR}$, i.e., $d_{TR} \geq 10$ m, the impact of $d_{p}$ becomes less significant since the capacity provided by VLC D2D is often surpassed by the capacity offered by RF D2D capacity. 

\begin{figure*}[t!]
\centering
\hspace{-1.3em}\begin{subfigure}[t]{0.365\textwidth}
\centering
\includegraphics[width=\textwidth]{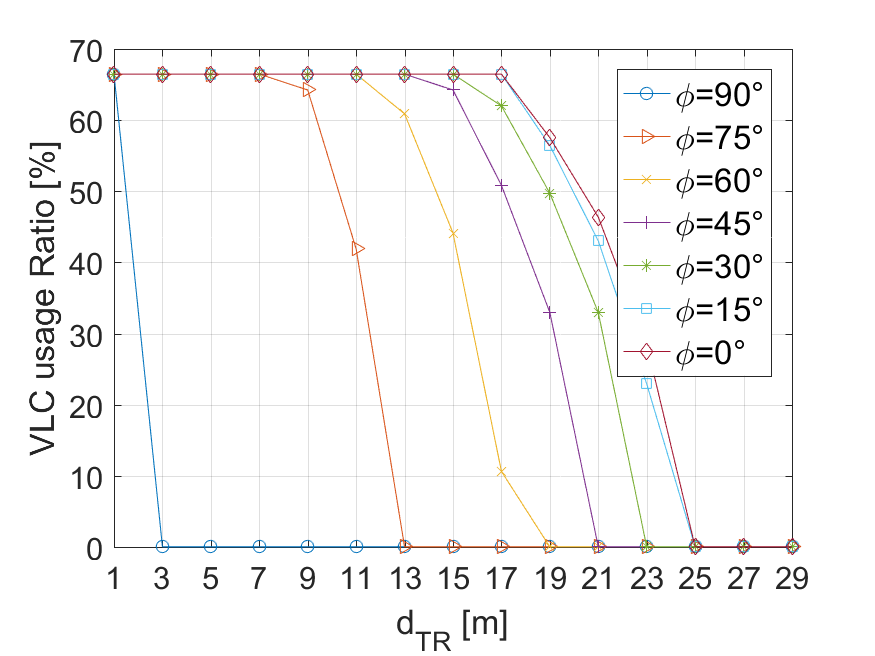}
\caption{$d_{P}=2$ m}
\end{subfigure}\hspace{-2.5em}
~
\begin{subfigure}[t]{0.365\textwidth}
\centering
\includegraphics[width=\textwidth]{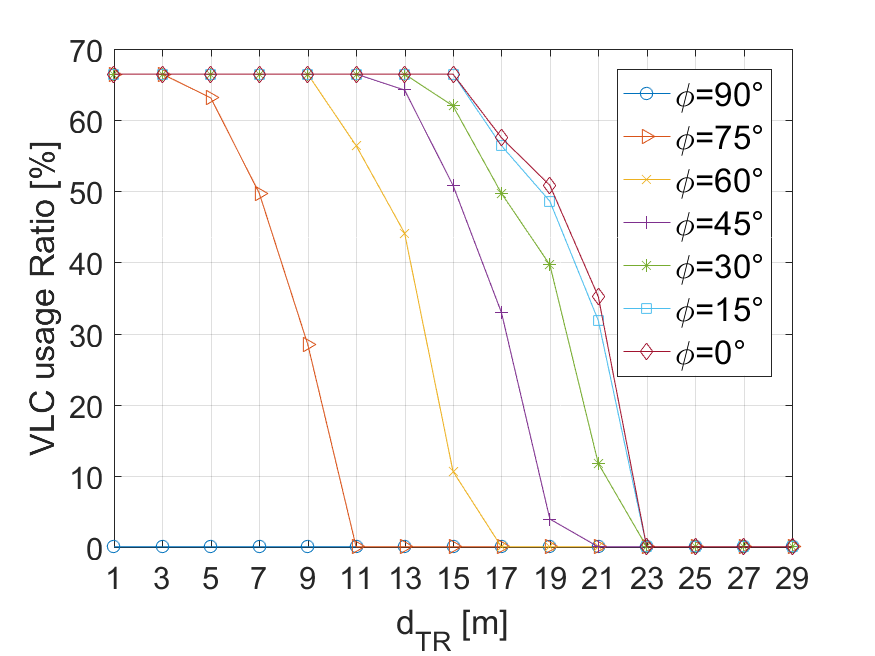}
\caption{$d_{P}=10$ m}
\end{subfigure}\hspace{-2.5em}
~
\begin{subfigure}[t]{0.365\textwidth}
\centering
\includegraphics[width=\textwidth]{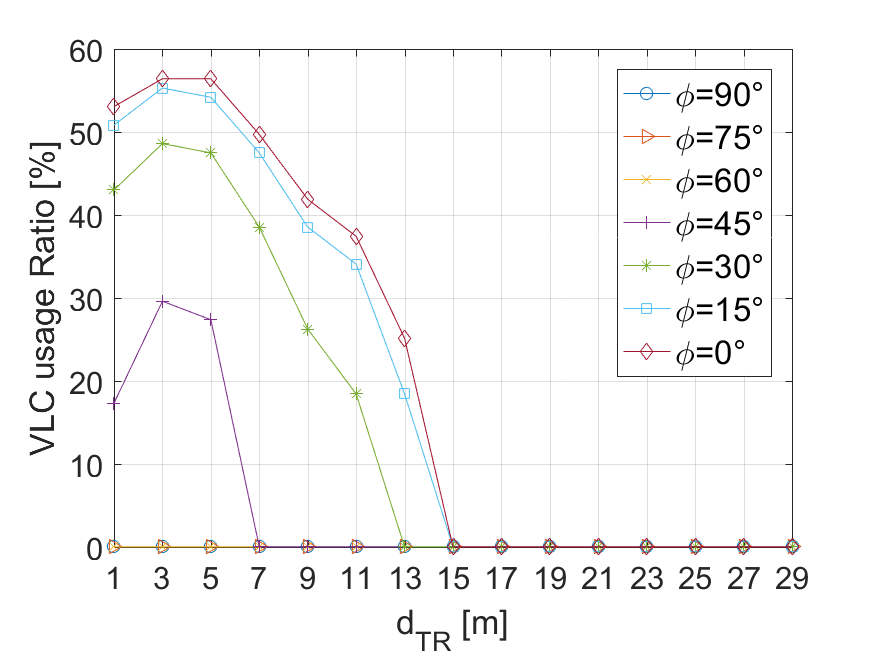}
\caption{$d_{P}=25$ m}
\end{subfigure}\hspace{-1.8em}
\caption{The ratio of time when VLC is used for communication instead of RF according to distance between transmitter and receiver ($d_{TR}$) and distance between pairs ($d_p$).}
\label{fig:2a}
\end{figure*}

From Fig.~\ref{fig:1a}, we can further see that the capacity raises as the orientation of MTs becomes close to the optimal (i.e., close to 0${^\circ}$). An important observation is that for $|\phi|\leq 30^{\circ}$ and $|\psi|\leq 30^{\circ}$, a degradation of the overall capacity is negligible (below 4\% with respect to the optimal angles). Even for $|\phi|\leq 60^{\circ}$ and $|\psi|\leq 60^{\circ}$, the degradation of capacity is still below 10\%. Furthermore, the impact of $\phi$ and $\psi$ is getting less important with rising $d_P$ because low interference in RF gives preference to a use of RF instead of VLC. The wide range of $\phi$ and $\psi$ angles that allow reaching almost the maximum capacity indicates that RF-VLC D2D can introduce significant gains even if the MTs are not directed towards each other. This observation is important for practical applications of the proposed concept as the orientation of both MTs is critical aspect in which the RF-VLC D2D concept differs from the common VLC communication assuming an access point located at the ceiling.

To understand better impact of both VLC and RF on the overall capacity of RF-VLC D2D, we analyze the ratio of time when VLC is used instead of RF. Fig.~\ref{fig:2a} shows that VLC mode is exploited in about 68\% if both $d_{TR}$ and $d_P$ are low. In this case, the capacity offered by VLC helps to improve the overall D2D performance and, thus, VLC is used predominantly. With increasing $d_{TR}$ and $d_{P}$, the orientation angle of MTs has to be closer to the optimal angle in order to keep VLC beneficial. If $d_{P}$ is equal or even longer than 25~m, VLC is not available and only RF mode can take place. In Fig.~\ref{fig:2a}c, we can also notice that for $d_{P} = 25$ m and $d_{TR} = 1$~m, VLC is used less often than for  $d_{TR} = 3$ or $5$~m. This is due to the fact that RF can perform very well if transmitter and receiver of the same pair are close to each other (i.e., low $d_{TR}$) while the interfering pair is far away (i.e., $d_{P}$ is high). The ratio of VLC usage confirms the fact that an indoor scenario with relatively close MTs is the most suitable for the proposed RF-VLC D2D concept as outlined in Section III.

\begin{figure*}[t!]
\centering
\hspace{-1.3em}\begin{subfigure}[t]{0.365\textwidth}
\centering
\includegraphics[width=\textwidth]{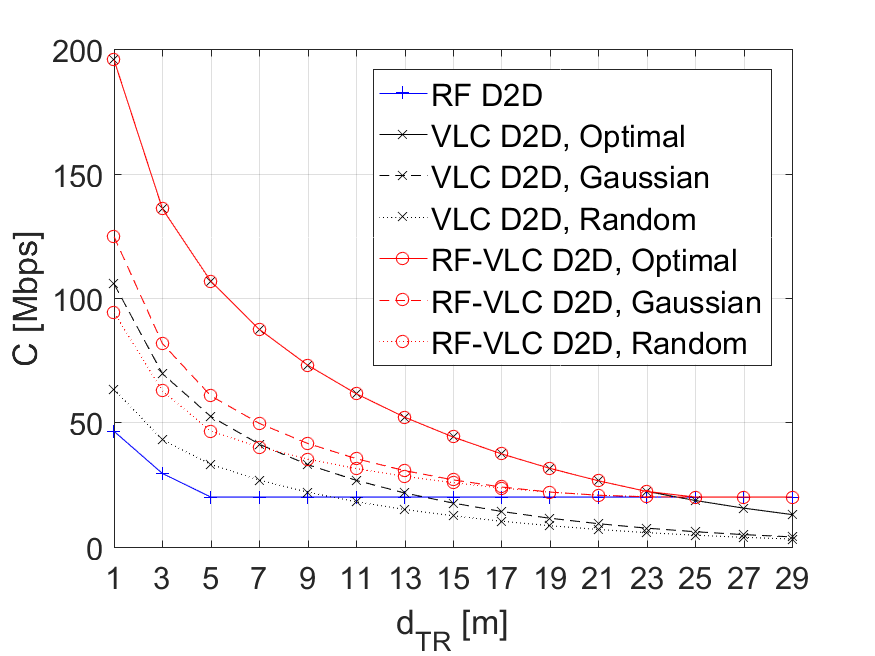}
\caption{$d_{P}=2$ m}
\end{subfigure}\hspace{-2.5em}
~
\begin{subfigure}[t]{0.365\textwidth}
\centering
\includegraphics[width=\textwidth]{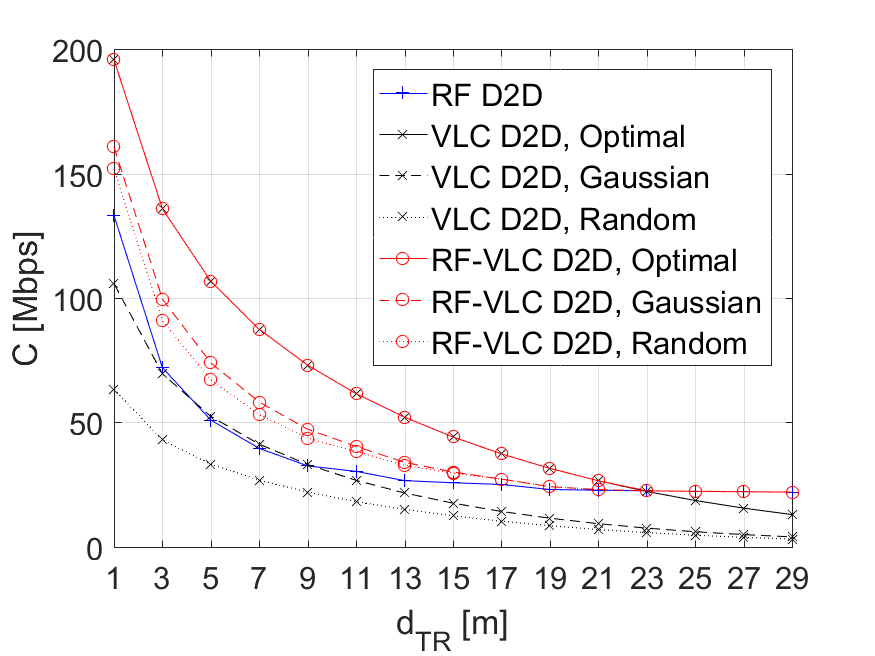}
\caption{$d_{P}=10$ m}
\end{subfigure}\hspace{-2.5em}
~
\begin{subfigure}[t]{0.365\textwidth}
\centering
\includegraphics[width=\textwidth]{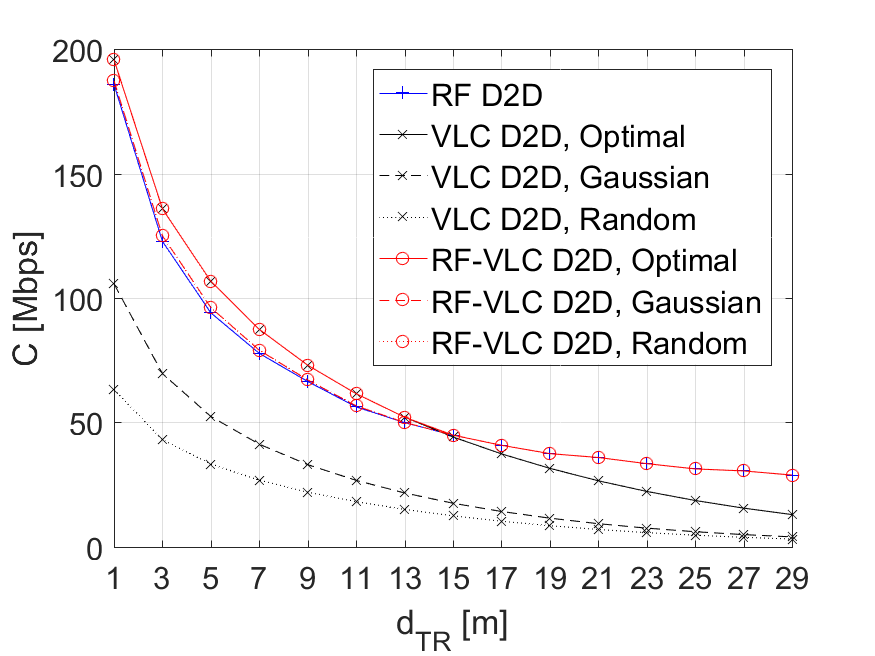}
\caption{$d_{P}=25$ m}
\end{subfigure}\hspace{-2em}
\caption{System capacity for RF D2D, VLC D2D, and RF-VLC D2D over distance between transmitter and receiver ($d_{TR}$) and distance between pairs ($d_p$). }
\label{fig:3a}
\end{figure*}

Now, we focus on comparison of the RF-VLC D2D with common RF D2D and VLC D2D systems as known today. We provide comparison for various $d_P$ in individual subplots of Fig.~\ref{fig:3a} and for three different ways of generation of $\phi$ and $\psi$ angles: Optimal, Gaussian, and Random, as described in Section IV.A (note that the results are averaged out over $10^6$ simulation drops). Fig.~\ref{fig:3a} shows that the proposed RF-VLC D2D outperforms both competitive schemes significantly and allows to provide maximum capacity disregarding $d_{TR}$ and $d_P$. More specifically, while RF D2D suffers in terms of capacity if $d_P$ is low, VLC D2D provides only limited capacity for high $d_{TR}$. In contrast, the proposed RF-VLC D2D performs well disregarding both distances. The most notable gain introduced by the novel RF-VLC D2D when compared to RF D2D is observed for low $d_{TR}$ and $d_P$, where RF-VLC D2D can provide 4.1, 2.6, and 2 times higher capacity for optimal, Gaussian, and Random selection of angles, respectively. At the same time, we can see that RF-VLC D2D outperforms VLC D2D even at short distances by 1.2 times (Gaussian selection of angles) and 1.5 times (Random selection). Note that for the Optimal selection of angles, VLC D2D and RF-VLC D2D perform similarly for low $d_{TR}$ because VLC is used in almost 100\% of time due to proximity of the $MT_T$ and $MT_R$. With increasing both $d_{TR}$ and $d_P$, the performance of the RF-VLC D2D converges to the conventional RF D2D since VLC is used only rarely. Moreover, with increasing $d_P$, the maximum $d_{TR}$ when VLC D2D still performs the same as the proposed RF-VLC D2D is decreasing. This is due to the fact that interference in RF is decreasing as well with increasing $d_P$ and consequently RF becomes more efficient.

\section{Conclusion}
In this paper, we have presented a novel D2D concept combining RF and VLC communication with the potential to increase the capacity provided by D2D. The performance analysis of the proposed RF-VLC D2D shows the ability to mitigate drawbacks in terms of limited capacity for very short and medium distances of the RF D2D and VLC D2D systems respectively. The proposed RF--VLC D2D increases the capacity by up to 4.1 and 1.5 times with respect to sole RF D2D and VLC D2D, respectively. The most notable gain in capacity is observed for low distances (up to 10 m), where VLC shows its superiority over conventional RF and, thus, the combination of both is the most beneficial.
 
The proposed RF-VLC opens many challenges needed to be addressed in the future. One of the key issues is an efficient selection of the communication mode, i.e., a decision when it is more profitable to exploit VLC, RF, or when to use both simultaneously in multi D2D pair scenario. Another challenge is also to understand better the changes of the MT's angles in a real world in order to develop machine learning techniques allowing to schedule data transmission to the most suitable band while reducing an amount of signaling required for the mode selection.

\section*{Acknowledgment}
This work has been supported by Grant No. P102/17/17538S funded by Czech Science Foundation.

\end{document}